\documentclass[10pt,conference]{IEEEtran}
\IEEEoverridecommandlockouts
\usepackage{cite}   
\usepackage{amsmath,amssymb,amsfonts}
\usepackage{graphicx}
\usepackage{textcomp}
\usepackage{tcolorbox}
\tcbuselibrary{listings,skins,raster}
\usepackage{algpseudocode}
\usepackage{listings}
\usepackage[T1]{fontenc}
\usepackage[utf8]{inputenc}
\usepackage{stmaryrd}
\usepackage{algorithm}
\usepackage{mathtools}
\usepackage{parcolumns}
\usepackage{booktabs}
\usepackage{xspace}
\usepackage{booktabs}
\usepackage[table,xcdraw]{xcolor}
\usepackage{colortbl}
\usepackage{cite}

\usepackage[colorlinks=true, linkcolor=blue, citecolor=blue, urlcolor=blue]{hyperref}
\colorlet{punct}{red!60!black}
\definecolor{background}{HTML}{EEEEEE}
\definecolor{delim}{RGB}{20,105,176}
\colorlet{numb}{magenta!60!black}
\definecolor{jsonkey}{rgb}{0.6,0,0}  
\definecolor{jsonvalue}{rgb}{0,0.3,0.6}
\def\BibTeX{{\rm B\kern-.05em{\sc i\kern-.025em b}\kern-.08em
    T\kern-.1667em\lower.7ex\hbox{E}\kern-.125emX}}
\lstdefinestyle{tightmono}{
  basicstyle=\ttfamily\fontsize{5.0}{5.5}\selectfont,
  columns=fullflexible,
  breaklines=true,
  breakatwhitespace=false,
  frame=tlrb,
  backgroundcolor=\color{background},
  aboveskip=1pt,
  belowskip=1pt,
  lineskip=-0.5pt,
  xleftmargin=0.5em,
  xrightmargin=0.5em,
  showstringspaces=false,
  keywordstyle=\color{blue},
  keywordstyle=[2]\color{red}\bfseries,
  morekeywords={POLICY,REQUESTS,FAULT,LOCALIZATION,
                Statement,stmt,Effect,Action,Resource,Expected,Responsible},
  morekeywords=[2]{V0,V1,V2,V3}
}

\begin{document}
\title{CloudFix: Automated Policy Repair for Cloud Access Control Policies Using Large Language Models}
\author{\IEEEauthorblockN{Bethel Hall}
\IEEEauthorblockA{\textit{Department of Computer Science} \\
\textit{Stevens Institute of Technology}\\
Hoboken, NJ, USA \\
bhall2@stevens.edu}
\and
\IEEEauthorblockN{Owen Ungaro}
\IEEEauthorblockA{\textit{Department of Computer Science} \\
\textit{Stevens Institute of Technology}\\
Hoboken, NJ, USA \\
oungaro@stevens.edu}
\and
\IEEEauthorblockN{William Eiers}
\IEEEauthorblockA{\textit{Department of Computer Science} \\
\textit{Stevens Institute of Technology}\\
Hoboken, NJ, USA \\
weiers@stevens.edu}
}
\maketitle
\begin{abstract}

Access control policies are vital for securing modern cloud computing, where organizations must manage access to sensitive data across thousands of users in distributed system settings. Cloud administrators typically write and update policies manually, which can be an error-prone and time-consuming process and can potentially lead to security vulnerabilities. Existing approaches based on symbolic analysis have demonstrated success in automated debugging and repairing access control policies; however, their generalizability is limited in the context of cloud-based access control. Conversely, Large Language Models (LLMs) have been utilized for automated program repair; however, their use for repairing such policies remains unexplored. In this work, we introduce CloudFix, the first automated policy repair framework for cloud access control that combines formal methods with LLMs. Given an access control policy and a specification of allowed and denied access requests, CloudFix employs formal methods-based Fault Localization to identify faulty statements in the policy and leverages LLMs to generate potential repairs, which are then formally verified using SMT solvers. To evaluate CloudFix, we curated a dataset of 282 real-world AWS access control policies extracted from forum posts and augmented them with synthetically generated request sets based on real scenarios. Our experimental results show that CloudFix improves repair accuracy over a baseline implementation across varying request sizes. Our work is the first to leverage LLMs for policy repair, showcasing the effectiveness of LLMs for access control and enabling efficient and automated repair of cloud access control policies. We make our tool and AWS dataset publicly available.
\end{abstract}


\section{Introduction}
Cloud compute systems allow administrators to write policies that specify access control rules, which govern how access to private data is granted. Policies are typically written in a convenient language, such as the Amazon Web Services (AWS) Identity and Access Management (IAM) policy language, allowing administrators to write their own policies. However, this is a tedious, complex, and often error-prone process. Moreover, policies must be modified on a regular basis due to the ever-changing nature of cloud computing. Creating and maintaining policies is a challenging endeavor, often requiring significant technical skills and domain expertise to ensure access control is correctly implemented. This can cause smaller organizations or those without the proper resources to forego best security practices in ensuring access privilege is kept to a minimum, and instead provide broad access to ensure their cloud computing services work as intended. 

Cloud computing services provide tools to assist administrators in writing and maintaining policies or monitoring access to resources (e.g., AWS IAM Policy Simulator, AWS CloudTrail service). However, when policies do not function as intended, or must be modified to incorporate access to new services or resources, these tools are inadequate when administrators lack the technical expertise. This is compounded by the lack of automated approaches for creating or repairing policies, as existing techniques either do not extend to cloud policies or have limited repair capabilities~\cite{Son2013FixMU,peng2017towards,D'Antoni2024,DetecingResolvingPolicyMisconfig,Jayaraman2014AutomatedAA}. What is needed is automated techniques that can reason about the complexities of cloud policies and can modify and repair policies as permissions and services evolve throughout the cloud ecosystem.

Recently, Large Language Models (LLMs) have seen great success in the field of software engineering~\cite{LLM4SESurvey}. They have become invaluable tools for software developers and engineers with their ability to generate code and automate tasks. In particular, LLMs are useful for automated program repair, where LLMs generate patches to fix faulty programs~\cite{zhang2024systematic,li2025hybrid,hossain2024deep,xia2023automated,campos2025empirical,orvalho2025counterexample}. In the field of access control, LLMs are successful in synthesizing policies from natural language descriptions~\cite{jayasundara2023sokaccesscontrolpolicy,vatsa2025synthesizingaccesscontrolpolicies,1269406,EnforcingAccessControlLLMS}. However, their applicability in the context of repairing access control policies has yet to be explored.

In this work, we take the first step towards leveraging Large Language Models for access control policy repair. Given a faulty or misconfigured access control policy and access control requests that must be allowed or denied by the policy, our approach repairs the policy so that it correctly classifies the given set of requests. We implement our approach in a tool called CloudFix and showcase its effectiveness in repairing AWS IAM access control policies. Additionally, we curate the first policy repair dataset for cloud access control policies and make it publicly available for future research endeavors. Specifically, we make the following key contributions: 
\begin{itemize}
\item We automate the access control policy repair process using an LLM-driven Fault Localization guided approach.
\item We present the first integration of LLMs with SMT-based formal verification for repairing access control policies;
\item We experimentally show that our Fault Localization techniques enable LLMs to repair faulty policies.
\item We release the public dataset for cloud policy repair and make the source code for CloudFix available on GitHub. 

\end{itemize}
The paper is organized as follows. In Section~\ref{sec:overview} we provide an overview of access control policies and the policy repair problem. In Section~\ref{sec:meth} we introduce CloudFix. In Section~\ref{sec:dataset}, we describe how we curate a dataset to evaluate CloudFix. In Section~\ref{sec:experiments} we experimentally validate CloudFix. In Section~\ref{sec:discussion}, we discuss our findings. In Section~\ref{sec:related} we discuss the related work. In Section~\ref{sec:conclusion}, we conclude our paper.

\section{Overview and Background}~\label{sec:overview}
In this work, we focus primarily on access control policies written in AWS IAM policy language. However, the techniques we present in this work are not specific to the AWS IAM policy language and can be applied to any policy language.

\subsection{Amazon Web Service Policies}
An AWS IAM policy $P$ is a declarative policy consisting of a list of statements describing how access is governed. Policies field access control requests, which are either allowed or denied by the policy. An access control request is a tuple $(\delta,a,r,e)$ where $\delta$ is a specific principal performing the access, $a$ is the action to be performed, $r$ is the resource being accessed, and $e$ is the set of environment variables within the request (e.g., IP address, current time, etc). A policy statement $S$ is a 5-tuple (\textit{Effect, Principal, Action, Resource, Condition}) where:
\begin{itemize}
    \item \textit{Effect} is either Allow or Deny
    \item \textit{Principal} consists of a list of users or entities
    \item \textit{Action} is a list of actions 
    \item \textit{Resource} is a list of resources 
    \item \textit{Condition} is an optional list of conditions further specifying how access is constrained
\end{itemize}
Initially, all requests are implicitly denied. An access control request is granted access if and only if there is at least one statement within the policy that explicitly allows access and no statement in the policy explicitly revokes or denies access. The AWS policy language allows special characters to be used which represent broad access: the '*' wildcard character can be used to represent any string, while the '?' character can be used to represent any character. Additionally AWS allows identity-based policies, which omit the Principal field.

\begin{figure*}[t]
\centering
  \includegraphics[width=0.9\linewidth]{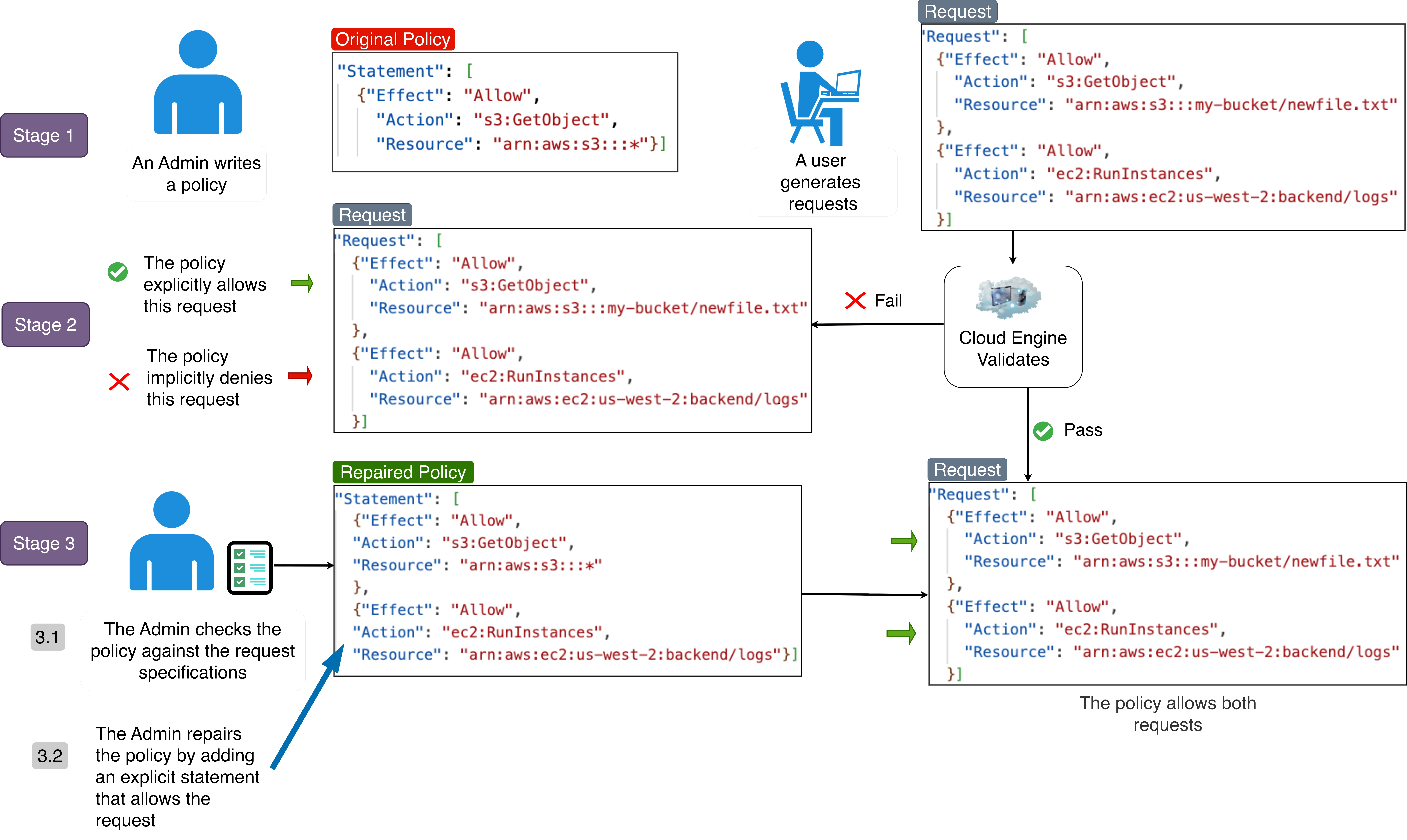}
  \caption{Manual Policy Repair Process. Stage 1: administrator authors policy, user generates requests. Stage 2: cloud engine validates requests, allowing s3:GetObject but implicitly denying ec2:RunInstances. Stage 3: The administrator adds an Allow statement and resubmits for validation.}
  \label{fig:manual}    
\end{figure*}    

\subsection{A Motivating Example}
In practice, debugging and repairing access control policies is largely manual. Administrators write policies in the AWS IAM console, then iteratively test them using the IAM Policy Simulator or CloudTrail logs until the desired permissions are achieved. A common workaround is to craft overly broad permissions (e.g., \texttt{Action: s3:"*"}) to avoid unintended denials, then narrow scope iteratively—a risky approach.

Figure~\ref{fig:manual} illustrates this process. In Stage~1, an administrator authors a policy granting \texttt{s3:GetObject} access. A user then submits two requests: one for \texttt{s3:GetObject} (allowed) and one for \texttt{ec2:RunInstances} (implicitly denied, as no statement allows it). In Stage~2, the Cloud Engine validates these requests and returns the misclassification. In Stage~3, the administrator manually adds an Allow statement for the denied request and redeploys.

This process is tedious and error-prone. Each misclassification requires inspecting the policy line by line, tracing evaluation logic, and cross-checking CloudTrail logs. As the number of users, policies, and requests grows, manual repair becomes increasingly impractical.

\section{CloudFix}~\label{sec:meth}
\begin{figure}[t]
\centering
  \includegraphics[width=0.8\linewidth]{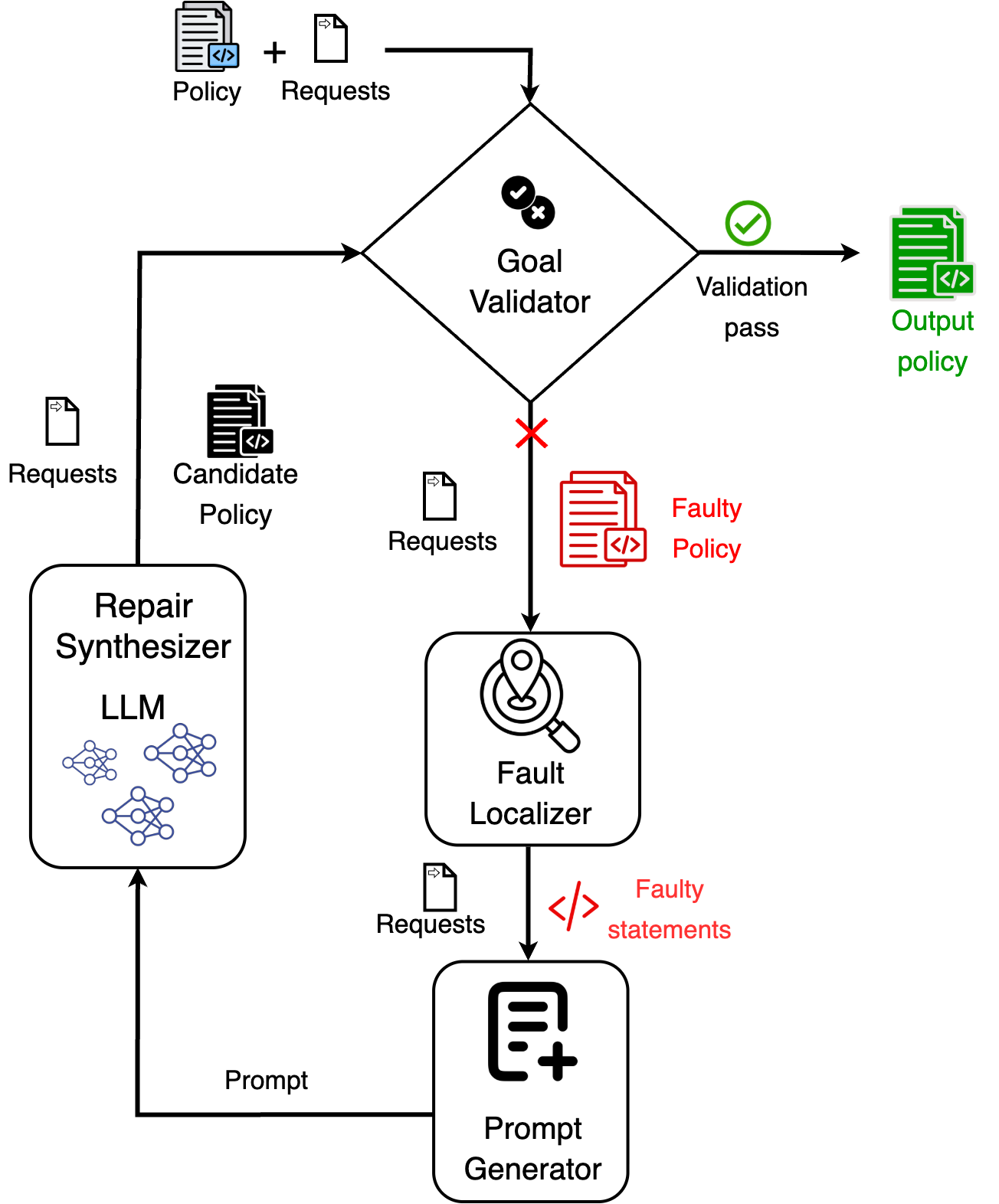}
  \caption{Architecture of CloudFix. 
  }
  \label{fig:method}    
\end{figure}

In this section, we introduce our policy repair framework, CloudFix.  It consists of four steps: (A) Goal Validation, (B) Fault Localization, (C) Prompt Generation, and (D) Repair Synthesis.  Figure~\ref{fig:method} shows the overall stages of the method. CloudFix operates over an input access control policy and request specification. The \textit{Goal Validator} first evaluates the policy against the request specification: if all requests are correctly classified, the policy is considered correct, and the framework returns the policy. Otherwise, the process proceeds iteratively. In each iteration, the faulty policy and the requests are fed into the \textit{Fault Localizer} component that analyzes the failed requests and the current policy to identify statements that are responsible for misclassifications. These faulty statements, together with the policy and requests, are provided to the \textit{Prompt Generator}, which produces an input prompt for the \textit{Repair Synthesizer}. The \textit{Repair Synthesizer }queries an LLM to generate a candidate policy as a potential repair. This candidate policy is validated against the requests by the \textit{Goal Validator}. If the candidate policy correctly classifies all requests, then the policy has been completely repaired, and the algorithm terminates. Otherwise, the candidate policy becomes the new input to the next iteration of the loop. This process continues until a fully repaired policy is produced, or the iteration threshold is reached. We retain the policy with the highest accuracy once the maximum iteration is reached. Alg.~\ref {alg:repair-highlevel} presents the complete CloudFix algorithm. 

\subsection{Goal Validator}
The \textit{Goal Validator} determines if the policy correctly classifies all requests in the given specification. Alg. \ref{alg:validate} shows how a request specification, $r$ in a set $R$ is validated against a policy $P$ to test whether $r$ is not allowed by $P$ and determine that $P$ is faulty. To do this, both the policy $P$ and $r$ are encoded into logical formulas suitable for satisfiability checking using the Quacky tool~\cite{eiers2022quacky}.
The algorithm returns FAIL if at least one request is misclassified. This information is then passed to the \textit{Fault Localizer} for further identification of statements within $P$ that are responsible for the misclassification of the $r$ in $S$. Encoding the requests and the policies as SMT formulas using an SMT Solver enables us to eliminate the dependency on external policy evaluation engines. We use a maximum of $I$ iterations per policy to reach a repair goal of 100\%. This can be defined by the user depending on the available resources. Once the target accuracy or the maximum $I$ is reached, the repair algorithm stops and returns the repaired policy. If the target repair accuracy is not achieved within $I$ iterations, the algorithm retains the repair with the highest accuracy. 
\begin{algorithm}[t]
\caption{\textsc{RepairPolicy}}
\label{alg:repair-highlevel}
\begin{algorithmic}[1]
\Require Policy $P$, Requests $R = \{ R_{\text{allow}}, R_{\text{deny}} \}$, Maximum iterations $I$
\Ensure Repaired Policy $P$ 
\State $verification \gets \textsc{ValidateGoal}(P, R)$
\If{$\mathit{verification} = \textsc{pass}$}
    \State \Return $P$ \Comment{Policy already satisfies requests}
\Else
    \For{$i = 1$ to $I$}
        \State $S_f \gets \textsc{LocalizeFaults}(P, R)$
        \State $prompt \gets \textsc{GeneratePrompt}(P, R, S_f)$
        \State $P^* \gets \textsc{SynthesisRepairWithLLM}(prompt)$
        \State $\mathit{verification} \gets \textsc{ValidateGoal}(P^*, R)$
        \If {$\mathit{verification} = \textsc{pass}$}
            \State \Return $P$
        \ElsIf{\text{P* gets better than P}}
            \State $P = P^*$
        \EndIf
    \EndFor
    \State \Return \textsc{P}
\EndIf
\end{algorithmic}
\end{algorithm}

\begin{algorithm}[t]
\caption{\textsc{ValidateGoal}}
\label{alg:validate}
\begin{algorithmic}[1]
\Require Policy $P$, Requests $R = \{ R_{\text{allow}}, R_{\text{deny}} \}$
\Ensure PASS or FAIL indicating whether $P$ is faulty
\State $\Phi \gets \textsc{Encode}(P)$
\For{each $(\delta,a,r,e) \in R_{\text{allow}}$ in $R$}
    \If{$(\delta,a,r,e) \not\models \Phi$}
        \State \Return \textsc{FAIL} \Comment{Missed allow}
    \EndIf
\EndFor
\For{each $(\delta,a,r,e) \in R_{\text{deny}}$ in $R$}
    \If{$(\delta,a,r,e) \models \Phi$}
        \State \Return \textsc{FAIL} \Comment{Missed deny}
    \EndIf
\EndFor
\State \Return \textsc{PASS} \Comment{Policy satisfies all goals}
\end{algorithmic}
\end{algorithm}
\subsection{Fault Localizer}
\label{sec:Fault Localization}
Once the \textit{Goal Validator} determines that a policy $\mathit{P}$  is faulty (i.e., it misclassifies requests), then $\mathit{P}$  is passed to the Fault Localizer to identify statements within it that misclassify the requests. 
We define a request misclassification as allowing a request that is supposed to be denied and/or denying a request that is supposed to be allowed by the policy. We define three types of faults within the policy that are responsible for misclassifications:
case 1) faulty statements that explicitly allow requests which should be denied as per the request specification, case 2) statements that explicitly deny requests that should be allowed per the request specification, and case 3) requests which are implicitly denied yet should be allowed.
The process is shown in Alg. \ref{alg:fault-loc-solver}. Each misclassified request is analyzed with respect to the policy to determine the type of fault that incorrectly classifies it. 

Given a policy $\mathit{P}$ and a request set $\mathit{R}$, lines 5 - 19 in Alg. \ref{alg:fault-loc-solver} illustrate how we identify faulty statements within $\mathit{P}$. The $\mathrm{SMTSolver}$ function determines the satisfiability of policy constraints against specific requests using the Z3 solver~\cite{z3} for satisfiability checking and the Quacky tool for policy and request encoding. 
For each misclassified request, to check if it falls into \textit{case~1}, $\mathrm{SMTSolver}(r,s)$ iteratively checks each allow statement $\mathit{s\in S^+}$ to see if it allows request $\mathit{r}$ by creating a minimal policy containing only that statement. Lines 8 and 9 show that if $\mathrm{SMTSolver}(r,s)$ returns $\mathit{SAT}$, we assert that there is an explicit allow statement that is allowing the statement that should have been denied. To check \textit{case~2}, $\mathrm{SMTSolver}(r, P^\star \cup \{s\}))$ combines the universal policy with a deny statement, $\mathit{s}$, to determine if the deny statement explicitly denies $\mathit{r}$. This is shown on line 11. If the solver returns $\mathit{UNSAT}$, it indicates that the deny statement in $\mathit{S}$ is responsible for the incorrect denial of $\mathit{r}$. If this condition fails, \textit{case~3} holds, as shown in line 14.  Finally, Alg. \ref{alg:fault-loc-solver} returns a list that has a mapping between policy statements and request failures, along with the failure types. This combination of statements, along with the explicit requests they allow, deny, or implicitly deny, is then passed to the \textit{Prompt Generator}, which is discussed below.

\subsection{Prompt Generator}
Once the \textit{Fault Localizer} identifies the faulty statements $s$ within $P$, we format the results into a structured natural language prompt for the LLM. The\textit{ Prompt Generator} receives the Fault Localization report, the list of requests $R$, and the policy $P$. It then structures them into a format suitable for LLMs. We annotate the prompts with the iteration number and the accuracy obtained from the \textit{Goal Validator} to provide contextual feedback to the LLM. 
In this work, we consider two types of prompts: baseline and Fault Localization guided.
\paragraph{\textbf{Baseline prompt}} This is a simple prompt we employ to simulate a real-world scenario where an engineer has access only to a policy and requests. It consists of the faulty policy that needs to be repaired in accordance with the must-allow and must-deny requests. 
\paragraph{\textbf{Fault Localization guided prompt}} We construct this prompt by first identifying the requests that each statement in a policy misclassifies. This augments the baseline prompt by providing the LLM with detailed information that maps specific failures to their root causes within the policy using an SMT-based Fault Localization report. Figure \ref{prompt} illustrates an example.  The prompt includes the policy, the must-allow and must-deny requests, and a Fault Localization report that indicates the error type, the failed requests, and the responsible statements for the failure. The complete prompts used in our experiments are provided in the accompanying artifact repository (see Section~\ref{sec:data}).

\begin{figure}[t]
\centering
\begin{minipage}{1.00\linewidth}
\begin{lstlisting}[style=tightmono]
POLICY: 
  Statement:{[
    {Sid: V1, Effect: Allow, Action: iam:GetAccountSummary, Resource: *}
    {Sid: V2, Effect: Allow, Action: s3:Get*, Resource: *}
    {Sid: V3, Effect: Deny,  Action: s3:PutObject, Resource: * }]}
\end{lstlisting}
\end{minipage}

\vspace{0.2em}
\noindent
\begin{minipage}[t]{0.50\linewidth}
\begin{lstlisting}[style=tightmono]
REQUESTS:

  Effect: Deny,
  Action: [s3:GetObject],
  Resource: [arn:aws:s3:::admin/doc1]


  Effect: Allow,
  Action: [s3:DeleteObject],
  Resource: [arn:aws:s3:::user/data]

  Effect: Deny,
  Action: [s3:DeleteObject],
  Resource: [arn:aws:s3:::admin/var]

  Effect: Deny,
  Action: [s3:ListBucket],
  Resource:[arn:aws:s3:::admin]


  Effect: Allow,
  Action: [s3:PutObject],
  Resource:[arn:aws:s3:::user/history]
    
\end{lstlisting}
\end{minipage}\hfill
\begin{minipage}[t]{0.49\linewidth}
\begin{lstlisting}[style=tightmono]

FAULT LOCALIZATION:

SHOULD BE DENIED BUT EXPLICITLY ALLOWED

Action:[s3:GetObject]
Resource:[arn:aws:s3:::admin/doc1]
Expected:deny, Got: allow
Responsible stmt: V2

SHOULD BE ALLOWED BUT EXPLICITLY DENIED

Action:[s3:PutObject]
Resource:[arn:aws:s3:::user/history]
Expected:allow, Got: deny
Responsible stmt: V3

SHOULD BE ALLOWED BUT IMPLICITLY DENIED

Action: [s3:DeleteObject]
Resource:[arn:aws:s3:::user/data1]
Expected:allow, Got: deny
Responsible stmt: None (missing allow)
  
\end{lstlisting}
\end{minipage}

\caption{Example of a Fault Localization guided prompt showing the faulty policy, requests, and fault localization report mapping failures to responsible statements.}
\label{prompt}
\end{figure}

\subsection{Repair Synthesizer (LLM)}
We leverage an LLM to generate candidate repairs directly from misclassified requests and Fault Localization findings.  The \textit{Repair Synthesizer LLM} produces a candidate policy, which is then passed to the \textit{Goal Validator} for verification. If the \textit{Goal Validator} determines that the policy correctly classified the request specification, the policy will be returned as a successful repair. However, due to the complexity of the requests and the policies themselves, some faulty policies are difficult to fully repair in a single iteration. The repair cycle continues until the \textit{Goal Validator} determines that the candidate policy correctly classifies the given must-allow and must-deny request specifications, or the predefined iteration threshold is reached. The candidate policy, which achieved the highest classification accuracy, is returned as a partially repaired policy.

\begin{algorithm}[t]
\caption{\textsc{LocalizeFaults}}
\label{alg:fault-loc-solver}
\begin{algorithmic}[1]
\Require Request set $R$, policy $P$ with statements $S$
\Ensure Fault list $L$
\State $S^{+}=\{s\in S \mid s.\mathrm{Effect}=\mathrm{Allow}\}$
\State $S^{-}=\{s\in S \mid s.\mathrm{Effect}=\mathrm{Deny}\}$
\State Let $P^\star$ be the universal-allow policy
\State $L \gets \emptyset$

\For{each $r \in R$}
  \If{$\textsc{ValidateGoal}(P, \{r\}) = \mathrm{FAIL}$}
    \If{$r.\mathrm{expected}=\mathrm{Deny}$ and $r.\mathrm{actual}=\mathrm{Allow}$}
      \State $A_r \gets \{\, s\in S^{+} \mid \mathrm{SMTSolver}(r,\{s\})=\mathrm{SAT}\,\}$
      \State $L \gets L \cup \{(r, A_r)\}$ \Comment{Case 1: explicit allow}
    \ElsIf{$r.\mathrm{expected}=\mathrm{Allow}$ and $r.\mathrm{actual}=\mathrm{Deny}$}
      \State $D_r \gets \{\, s\in S^{-} \mid \mathrm{SMTSolver}(r, P^\star \cup \{s\})=\mathrm{UNSAT}\,\}$
      \If{$D_r \neq \emptyset$}
        \State $L \gets L \cup \{(r, D_r)\}$ \Comment{Case 2: explicit deny}
      \Else
        \State $L \gets L \cup \{(r, \emptyset)\}$ \Comment{Case 3: implicit deny}
      \EndIf
    \EndIf
  \EndIf
\EndFor
\State \Return $L$
\end{algorithmic}
\end{algorithm}

\section{Dataset Collection and Request Generation}~\label{sec:dataset}
CloudFix works to repair cloud access control policies that are written in cloud access control languages. However, there is currently no public dataset specifically curated for  \textit{policy repair}, which makes research in this area difficult. This is because policies model access control within organizations, and if these policies were made public, they could create security vulnerabilities. To address this gap, we curate real-world AWS IAM policies from the \href{https://repost.aws/questions}{AWS re:Post} forum and make them available to the research community. An important component of policy repair is the availability of representative request sets, but obtaining such a dataset is non-trivial because AWS IAM request logs are also kept private. To overcome these gaps, we algorithmically generate synthetic requests directly from the structure of the policy by varying the proportions of must-allow and must-deny requests.
\subsection{Policy Collection}
The policy collection process involves three steps: first, scraping policies from public sources, then detecting and categorizing valid JSON policies, and finally normalizing and repairing formatting inconsistencies. To collect a sufficiently large dataset of IAM policies, we sourced user posts from \href{https://repost.aws/questions}{AWS re:Post}. In this community forum, AWS users share questions and answers regarding AWS services and 
products. We specifically targeted posts containing IAM policies that either failed to function as intended or required correction. 
We scraped 9,968 IAM-related posts and scanned each for JSON structures containing IAM policy keys such as Effect, Action, or Statement. We validated all collected policies using Quacky\cite{eiers2022quacky}, a quantitative policy analyzer, which revealed widespread structural errors and formatting inconsistencies. After applying a naive syntactic repair tool to enforce standard IAM conventions, we obtained 282 faulty policies (syntactically valid but semantically broken) and 45 corresponding community-accepted repairs. Table~\ref{tab:dataset-stats} summarizes our dataset.

\definecolor{headerblue}{RGB}{68, 114, 196}
\definecolor{lightblue}{RGB}{227, 236, 248}
\definecolor{wingreen}{RGB}{34, 139, 34}
\begin{table}[t]
\centering
\caption{Summary Statistics of the AWS IAM Policy Dataset}
\label{tab:dataset-stats}
\small
\setlength{\tabcolsep}{8pt}
\renewcommand{\arraystretch}{1.25}
\begin{tabular}{l r}
\toprule
\rowcolor{headerblue}
\textcolor{white}{\textbf{Metric}} & \textcolor{white}{\textbf{Value}} \\
\midrule
\rowcolor{lightblue} Total Policies & 282 \\
Total Statements & 444 \\
\rowcolor{lightblue} Avg. Statements per Policy & 1.58 \\
Min Statements per Policy & 1 \\
\rowcolor{lightblue} Max Statements per Policy & 10 \\
Unique Services & 92 \\
\rowcolor{lightblue} Unique Actions & 664 \\
Unique Resource Types & 189 \\
\rowcolor{lightblue} Cross-Service Policies & 51 \\
\midrule
\rowcolor{headerblue}
\multicolumn{2}{l}{\textcolor{white}{\textbf{Effect Distribution}}} \\
\rowcolor{lightblue} \quad Allow & 400 (89.7\%) \\
\quad Deny & 44 (9.9\%) \\
\bottomrule
\multicolumn{2}{l}{\scriptsize\textit{Percentages are relative to total statements.}} \\
\end{tabular}
\end{table}

\subsection{Request Generation}
Alg. \ref{alg:policy-request-generator} shows our approach to automatically generating synthetic requests to evaluate our policy repair framework. We begin by extracting each element of the policy and then generate requests, making them 60\% allowed and 40\% denied. We include all elements of the policy (i.e., effect, action, resource, condition, and principal). To create a realistic repair, we add misclassified requests by selecting a subset of the truly allowed and/or denied requests and modifying either of the values in the policy keys to invert the expected effect (i.e., allow or deny). The SampleAllowed$(\mathcal{E},\lfloor 0.6n\rfloor)$ function creates approximately 60\% requests expected to be allowed, while \textsc{SampleDenied}$(\mathcal{E}, n-|\mathcal{R}^+|)$ (line 24) generates the remaining 40\% expected to be denied. Note that 89.7\% of statements in the policy dataset are \textit{allow} statements (see Table~\ref {tab:dataset-stats}), and the 60\%-40\% split mentioned here applies only to the synthetically generated requests. We combine these two request sets in line 5. In line 6, \textsc{RandomSubset}$(\mathcal{R},\lfloor\rho \cdot n\rfloor)$ selects a fraction $\rho$ of the total requests to be intentionally misclassified by altering the principal, conditions, and resources (e.g., changing the regions in the resources). This ensures that the final dataset contains both correctly and incorrectly classified examples for a robust evaluation of the method and allows us to parameterize the misclassification rate in order to evaluate our approach. From lines 24 - 31, \textsc{SampleDenied} generates requests by sampling a combination of policy elements that are expected to be denied by the policy. These requests are different from those in the allowed requests in at least one element. Finally, \textsc{GenerateRequests} returns the set of allowed, denied, and misclassified requests. 
\begin{algorithm}[t]
\caption{\textsc{GenerateRequests}}
\label{alg:policy-request-generator}
\begin{algorithmic}[1]
\Function{GenerateRequests}{$\mathcal{P}$, $n$, $\rho$}
   \State $\mathcal{E} \gets$ \Call{Extract}{$\mathcal{P}$} \Comment{Extract policy elements}
   \State $\mathcal{R}^+ \gets$ \Call{SampleAllowed}{$\mathcal{E}$, $\lfloor 0.6n \rfloor$}
   \State $\mathcal{R}^- \gets$ \Call{SampleDenied}{$\mathcal{E}$, $n - |\mathcal{R}^+|$}
   \State $\mathcal{R} \gets \mathcal{R}^+ \cup \mathcal{R}^-$
   \State $\mathcal{M} \gets$ \Call{RandomSubset}{$\mathcal{R}$, $\lfloor \rho \cdot n \rfloor$} \Comment{Select subset to misclassify}
   \For{each $r \in \mathcal{M}$}
       \State $r.\text{Effect} \gets$ \Call{Flip}{$r.\text{Effect}$} \Comment{Change allow to deny or vice versa}
       \State $r \gets$ \Call{ModifyAttributes}{$r$} \Comment{Change one of $(s,p,c)$}
   \EndFor
   \State \Return {$\mathcal{R}$}
\EndFunction

\Function{SampleAllowed}{$\mathcal{E}$, $k$}
   \State $\mathcal{R}^+ \gets \emptyset$
   \For{$i \gets 1$ \textbf{to} $k$}
       \State $a \gets$ \Call{Sample}{$\mathcal{E}.\text{actions}$}
       \State $s \gets$ \Call{Sample}{$\mathcal{E}.\text{resources}$}
       \State $p \gets$ \Call{Sample}{$\mathcal{E}.\text{principals}$}
       \State $c \gets$ \Call{Sample}{$\mathcal{E}.\text{conditions}$}
       \State $\mathcal{R}^+ \gets \mathcal{R}^+ \cup \{\langle a, s, p, c, \text{allow} \rangle\}$
   \EndFor
   \State \Return $\mathcal{R}^+$
\EndFunction

\Function{SampleDenied}{$\mathcal{E}$, $k$}
   \State $\mathcal{D} \gets$ \Call{GenerateComplements}{$\mathcal{E}$}
   \State $\mathcal{R}^- \gets \emptyset$
   \For{$i \gets 1$ \textbf{to} $k$}
       \State $(a, s, p, c) \gets$ \Call{Sample}{$\mathcal{D}$}
       \State $\mathcal{R}^- \gets \mathcal{R}^- \cup \{\langle a, s, p, c, \text{deny} \rangle\}$
   \EndFor
   \State \Return $\mathcal{R}^-$
\EndFunction
\end{algorithmic}
\end{algorithm}

\section{Experimental Design}~\label{sec:experiments}
We address the following research questions:\\
\noindent\textbf{RQ1: How effective is Fault Localization at improving LLM-based policy repair compared to a baseline?} 
\\
\noindent\textbf{RQ2: How does request size affect repair accuracy?} \\
\noindent\textbf{RQ3: How does repair accuracy vary across LLMs?} \\
\noindent\textbf{RQ4: Can CloudFix generate repairs that generalize to semantically similar requests beyond those provided during the repair process?} \\
\noindent\textbf{RQ5: Can CloudFix repair policies when request specifications are synthesized from natural language descriptions of user intent?}

\subsection{Experiment Setup}
All experiments were run using a Rocky Linux 9.3 computer cluster (Blue Onyx). Each job was allocated a single NVIDIA L40S GPU with 46\,GB of memory, two Intel(R) Xeon(R) Platinum 8562Y+ CPU cores, and 256\,GB of system RAM. All experiments were executed with a 64\,GB memory limit and a single GPU per task.

\subsection{Models} We evaluate our approach using four open-source instruction-tuned models of comparable size. Specifically, we used \textit{CodeLLaMA-7B-Instruct}, a model trained for code synthesis and reasoning; \textit{Granite-3.3-8B-Instruct}, an IBM model optimized for instruction following in software engineering tasks; \textit{DeepSeek-Coder-7B-Instruct}, a model finetuned for code completion and reasoning, designed for software repair and program synthesis, and \textit{Llama3.2-3B-Instruct}, an instruction-tuned text-only model that supports a maximum of 128K context window. We maintain similar prompts and model hyperparameters for all models used. All models were loaded and run via Hugging Face's transformers pipeline \cite{wolf2019huggingface}.

\subsection{Repair Approaches} We employ two approaches: first, a baseline repair approach that uses the basic prompt to instruct the LLM to repair a faulty policy. The baseline receives the complete faulty policy, the requests that must be allowed and denied, and a generic repair instruction, where the model needs to identify faulty statements within the policy and repair them. Second, a Fault Localization guided approach, described in detail in Section~\ref{sec:meth}, which represents the core contribution of CloudFix. The baseline serves as an ablation to isolate the impact of Fault Localization. We maintain the same user and system prompts for both approaches to ensure that the quality of the prompts does not confound the results.
Request size is another key factor that influences repair accuracy. Using Alg.~\ref{alg:policy-request-generator}, we generate request sets of size \texttt{10}, \texttt{20}, \texttt{30}, and \texttt{50}, where each request targets a unique resource–action–condition–principal combination to increase complexity.

\subsection{\textbf{Evaluation Metric}}
 \textbf{Accuracy.} We evaluate our approach using accuracy, which is defined as the percentage of request specifications that are correctly classified by the policy. $\text{Accuracy} = \frac{R_c}{R_{\text{total}}} \times 100\%$,
where $R_c$ is the number of correctly classified requests and $R_{\text{total}}$  is the total number of requests. We define three accuracy categories: \textit{$100\%$ (complete repair)}, \textit{$80-99\%$ (moderate repair)}, and \textit{$<80\%$ (failed repair)}. 

\textbf{Iteration, $I$.} We evaluate the total number of iterations the repair takes to reach a \textit{complete}, \textit{moderate}, or \textit{failed repair}. In our experiments, $I$ is set to 5.

\textbf{Time.} We measure the total repair duration, including the time that an LLM takes to generate a repair and the time the SMT solver takes to validate the repair across all iterations until either 100\% is achieved or the max $I$ is reached. 

\subsection{Experiments}

\textbf{RQ1: How effective is Fault Localization at improving LLM-based policy repair compared to a baseline? }Our first research question evaluates whether Fault Localization guidance improves the ability of LLMs to repair faulty access control policies compared to a baseline approach.

\textbf{Setup.} 
To answer RQ1, we evaluate our framework on 282 AWS IAM policies, generating \texttt{10}, \texttt{20},\texttt{30}, and \texttt{50} must-allow and must-deny requests per policy using Alg. \ref{alg:policy-request-generator}.

\textbf{Results.} Table \ref{tab:binned-fl-base} shows that Fault Localization achieves  84.0\% full repair rate on 10 requests compared to 
48.2\% for baseline, and maintains 54.3\% complete repair on 30 requests while baseline drops to 22.3\%. We see a performance degradation for 50 request sizes in both baseline and Fault Localization, where the baseline repairs 0 policies completely, while Fault Localization generates 7 completely repaired policies. The results demonstrate that Fault Localization consistently outperforms the baseline as the request size increases, and gracefully degrades as baseline performance dramatically decreases for higher request sizes.

We further performed a statistical significance analysis on the obtained results using a two-tailed \textit{t}-test to compare the mean repair accuracies between the two approaches. The p-values below 0.001 for request sizes 10, 20, and 30 show strong statistical evidence (>99.9\% confidence, see Table \ref{tab:significance}, shaded rows) that Fault Localization outperforms the baseline. However, the p-value of 0.056 for request size 50 fails to meet the significance threshold of 0.05, meaning the 3.85\% improvement could be due to random chance rather than the true impact of Fault Localization. 

\textbf{RQ2: How does request size affect repair accuracy?}

Our second research question examines how repair accuracy changes as the number of requests increases. We compare our two approaches: (1) baseline, which uses a simple prompt consisting of only the faulty policy and requests, and (2) Fault Localization-guided repair, which additionally provides the Fault Localization information discussed in Section~\ref{sec:meth}.

\textbf{Setup.} To answer RQ2, we use the same experimental setup as RQ1, evaluating both approaches on the same 282 policies with request sizes of \texttt{10}, \texttt{20}, \texttt{30}, and \texttt{50}. Both approaches use similar LLM configuration, prompt structure, and repair iteration limits (we set $I$ to 5 iterations). 

\textbf{Results.} Table~\ref{tab:binned-fl-base} shows the distribution of repair accuracy across different request sizes. The Fault Localization guided repair demonstrates significant improvements over the baseline for request sizes 10, 20, 30, and 50.  For request size 10, Fault Localization generates 237 complete repairs compared to only 136 for the baseline. This advantage continues with Fault Localization, achieving 170 complete repairs for 20 requests, 153 complete repairs for 30 requests, and 7 complete repairs for 50 requests. 

For request size 10, the baseline generates 101 policies in the $<$80\% category, while Fault Localization reduces this to only 19 policies. Similarly, for 30 requests, the baseline generates 100 policies with $<$80\% accuracy compared to just 30 with Fault Localization. However, the baseline achieves no 100\% repairs but only 67 policies in the 80--99\% range, while Fault Localization only achieves 7 complete repairs and produces 175 policies with $<$80\% accuracy compared to the 215 of the baseline. We observed that as the request size increases, the repair task may exceed the LLM's capability to integrate Fault Localization information effectively. The accuracy degradation at larger request sizes indicates that, while Fault Localization is effective for moderately complex repair tasks, additional techniques may be needed to maintain its benefits when the number of request sizes grows substantially.

\definecolor{sigblue}{RGB}{227, 236, 248}  
\definecolor{headerblue}{RGB}{68, 114, 196}  

\begin{table}[t]
\centering
\caption{Statistical Significance: Repair Accuracy of Baseline and Fault Localization}
\label{tab:significance}
\renewcommand{\arraystretch}{1.2}
\resizebox{\columnwidth}{!}{%
\begin{tabular}{ccccc}
\toprule
\rowcolor{headerblue}
\textcolor{white}{\textbf{Request Size}} & \textcolor{white}{\textbf{Baseline}} & \textcolor{white}{\textbf{FL}} & \textcolor{white}{\textbf{Difference}} & \textcolor{white}{\textbf{p-value}} \\
\rowcolor{headerblue}
& \textcolor{white}{(Accuracy \%)} & \textcolor{white}{(Accuracy \%)} & \textcolor{white}{($\Delta$ pp)} & \\
\midrule
\rowcolor{sigblue} 10 & 77.66 & \textbf{95.21} & +17.55 & $<$0.001\textsuperscript{***} \\
\rowcolor{sigblue} 20 & 83.40 & \textbf{90.69} & +7.29 & $<$0.001\textsuperscript{***} \\
\rowcolor{sigblue} 30 & 75.63 & \textbf{92.29} & +16.67 & $<$0.001\textsuperscript{***} \\
50 & 62.44 & 66.28 & +3.85 & 0.056 \\
\bottomrule
\multicolumn{5}{l}{\textsuperscript{***}$p < 0.001$. Shaded rows indicate statistical significance.}
\end{tabular}%
}
\end{table}

\begin{figure*}[t]
  \centering
  \includegraphics[width=0.9\linewidth]{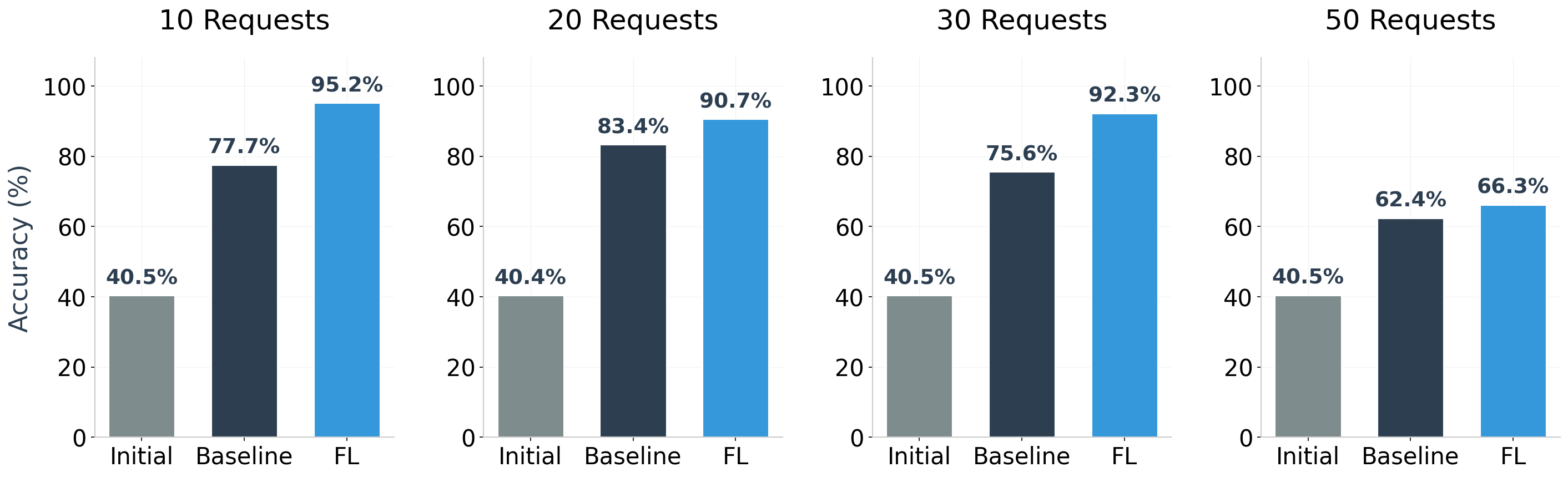}
  \caption{Average repair accuracy of Baseline and FL approaches across varying request sizes.}
  \label{fig:result}
\end{figure*}

\begin{table*}[t]
\centering
\small
\caption{Policy Repair Distribution by Request Size}
\label{tab:binned-fl-base}
\setlength{\tabcolsep}{6pt}
\renewcommand{\arraystretch}{1.25}
\begin{tabular}{l cc cc cc cc}
\toprule
\rowcolor{headerblue}
& \multicolumn{8}{c}{\textcolor{white}{\textbf{Request Size}}} \\
\rowcolor{headerblue}
\textcolor{white}{\textbf{Category}} & \multicolumn{2}{c}{\textcolor{white}{\textbf{10}}} & \multicolumn{2}{c}{\textcolor{white}{\textbf{20}}} & \multicolumn{2}{c}{\textcolor{white}{\textbf{30}}} & \multicolumn{2}{c}{\textcolor{white}{\textbf{50}}} \\
\rowcolor{headerblue}
& \textcolor{white}{Baseline} & \textcolor{white}{FL} & \textcolor{white}{Baseline} & \textcolor{white}{FL} & \textcolor{white}{Baseline} & \textcolor{white}{FL} & \textcolor{white}{Baseline} & \textcolor{white}{FL} \\
\midrule
\rowcolor{lightblue} 100\% Fix & 136 (48.2\%) & \textbf{237 (84.0\%)} & 149 (52.8\%) & \textbf{170 (60.3\%)} & 63 (22.3\%) & \textbf{153 (54.3\%)} & 0 (0.0\%) & \textbf{7 (2.5\%)} \\
80--99\% & 45 (16.0\%) & 26 (9.2\%) & 62 (22.0\%) & 69 (24.5\%) & 119 (42.2\%) & 99 (35.1\%) & 67 (23.8\%) & 100 (35.5\%) \\
$<$80\% & 101 (35.8\%) & 19 (6.7\%) & 71 (25.2\%) & 43 (15.3\%) & 100 (35.5\%) & 30 (10.6\%) & 215 (76.2\%) & 175 (62.1\%) \\
\bottomrule
\multicolumn{9}{l}{\scriptsize FL denotes Fault Localization; \textbf{bold} indicates the best complete repair rate for each request size.}

\end{tabular}%
\end{table*}

\definecolor{headerblue}{RGB}{68, 114, 196}
\definecolor{lightblue}{RGB}{227, 236, 248}
\definecolor{wingreen}{RGB}{34, 139, 34}

\begin{table*}[t]
\centering
\small
\caption{Repair Time for Baseline and Fault Localization Across Request Sizes}
\label{tab:timing_transposed}
\setlength{\tabcolsep}{6pt}
\renewcommand{\arraystretch}{1.25}
\begin{tabular}{l cc cc cc cc}
\toprule
\rowcolor{headerblue}
& \multicolumn{8}{c}{\textcolor{white}{\textbf{Request Size}}} \\
\rowcolor{headerblue}
\textcolor{white}{\textbf{Metric}} & \multicolumn{2}{c}{\textcolor{white}{\textbf{10}}} & \multicolumn{2}{c}{\textcolor{white}{\textbf{20}}} & \multicolumn{2}{c}{\textcolor{white}{\textbf{30}}} & \multicolumn{2}{c}{\textcolor{white}{\textbf{50}}} \\
\rowcolor{headerblue}
& \textcolor{white}{Baseline} & \textcolor{white}{FL} & \textcolor{white}{Baseline} & \textcolor{white}{FL} & \textcolor{white}{Baseline} & \textcolor{white}{FL} & \textcolor{white}{Baseline} & \textcolor{white}{FL} \\
\midrule
Avg. Iterations & 3.14 & \cellcolor{lightblue}\textbf{1.80} & 2.93 & \cellcolor{lightblue}\textbf{2.85} & 4.13 & \cellcolor{lightblue}\textbf{3.08} & 5.00 & \cellcolor{lightblue}\textbf{4.92} \\
Avg. Total Time (s) & 370.3 & \cellcolor{lightblue}\textbf{136.5} & 667.2 & \cellcolor{lightblue}\textbf{456.3} & 827.1 & \cellcolor{lightblue}\textbf{547.3} & 1778.6 & \cellcolor{lightblue}\textbf{1306.9} \\
Avg. SMT Time (s) & 2.98 & \cellcolor{lightblue}\textbf{0.27} & 8.58 & \cellcolor{lightblue}\textbf{0.44} & 11.05 & \cellcolor{lightblue}\textbf{1.11} & 8.26 & \cellcolor{lightblue}\textbf{3.92} \\
Avg. LLM Time (s) & 367.36 & \cellcolor{lightblue}\textbf{136.20} & 658.66 & \cellcolor{lightblue}\textbf{455.86} & 816.02 & \cellcolor{lightblue}\textbf{546.20} & 1770.29 & \cellcolor{lightblue}\textbf{1302.98} \\
\midrule
SMT Time \% & 0.80 & \cellcolor{lightblue}0.20 & 1.30 & \cellcolor{lightblue}0.10 & 1.30 & \cellcolor{lightblue}0.20 & 0.50 & \cellcolor{lightblue}0.30 \\
LLM Time \% & 99.20 & \cellcolor{lightblue}99.80 & 98.70 & \cellcolor{lightblue}99.90 & 98.70 & \cellcolor{lightblue}99.80 & 99.50 & \cellcolor{lightblue}99.70 \\
\bottomrule
\multicolumn{9}{l}{\scriptsize FL denotes Fault Localization; \textbf{bold} indicates faster convergence (fewer iterations).}
\end{tabular}%
\end{table*}


\textbf{RQ3: How does repair accuracy vary across different open-source LLMs?}
We now compare LLM performance on policy repair.

\textbf{Setup.} We employed the instruction-tuned variants of \textit{CodeLlama-7B-Instruct}, \textit{Llama-3.2-3B-Instruct}, \textit{DeepSeek-Coder-7B-Instruct-v1.5}, and \textit{Granite-3.3-8B-Instruct}. We prompted all models using the Fault Localization guided prompt with a request size of 30 for all 282 access control policies.

\begin{figure}[t]
  \centering
  \includegraphics[width=\linewidth]{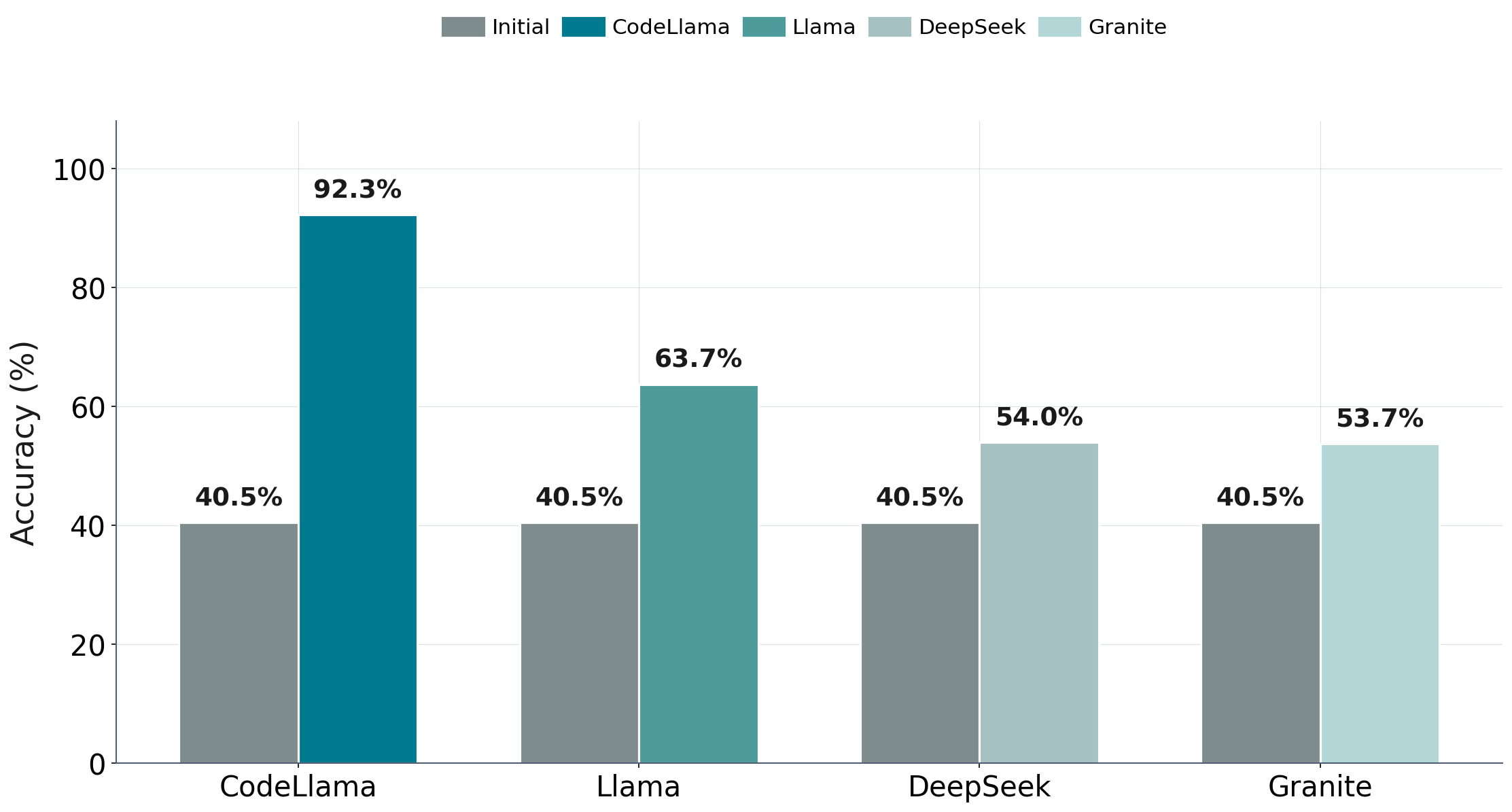}
  \caption{Overall repair accuracy of four open-source LLMs using Fault Localization guided with a request size of 30.}
  \label{fig:llms}
\end{figure}

\textbf{Results.} Figure \ref{fig:llms} shows the overall repair accuracy of each model. Notably, CodeLlama achieves the highest overall accuracy of 92.3\%, followed by Llama of 63.7\%. DeepSeek-Coder and Granite had a relatively similar overall repair accuracy of 54.0\% and 53.7\%. Table \ref{tab:fl_llm_comparison}  further shows the accuracy breakdown. CodeLlama was able to repair 54.3\% (complete repair rate) of the 282 policies . The low accuracy in the rest of the models is best explained in the third category, which shows an increasing number of failed repairs, going from left to right. 

\begin{table*}[t]
\centering
\caption{Fault Localization Repair Accuracy Comparison Across Different LLMs (Request Size 30)}
\label{tab:fl_llm_comparison}
\small
\setlength{\tabcolsep}{1pt}
\renewcommand{\arraystretch}{1.25}
\begin{tabular*}{\textwidth}{@{\extracolsep{\fill}}lcccc@{}}
\toprule
\rowcolor{headerblue}
\textcolor{white}{\textbf{Category}} & \textcolor{white}{\textbf{CodeLlama}} & \textcolor{white}{\textbf{Llama}} & \textcolor{white}{\textbf{DeepSeek-Coder}} & \textcolor{white}{\textbf{Granite}} \\
\midrule
100\% Fix & \cellcolor{lightblue}\textbf{153 (54.3\%)} & 19 (6.7\%) & 37 (13.1\%) & 39 (13.8\%) \\
80--99\% & \cellcolor{lightblue}\textbf{99 (35.1\%)} & 55 (19.5\%) & 32 (11.3\%) & 29 (10.3\%) \\
$<$80\% & \cellcolor{lightblue}\textbf{30 (10.6\%)} & 208 (73.8\%) & 213 (75.5\%) & 214 (75.9\%) \\
\bottomrule
\multicolumn{5}{l}{\scriptsize \textbf{Bold, shaded cells} denote the best-performing model (CodeLlama); results use Fault Localization prompting.}
\end{tabular*}
\end{table*}

\textbf{RQ4: Can CloudFix generate repairs that generalize to semantically similar requests beyond those provided during the repair process?} This question tests whether repaired policies capture general access patterns rather than overfitting to specific requests. For example, requests targeting 
\noindent\texttt{(s3:getObject,arn:aws:s3:::category/\textbf{data1})} \\
\noindent\texttt{(s3:getObject,arn:aws:s3:::category/\textbf{data2})} \\
\noindent\texttt{(s3:getObject,arn:aws:s3:::category/\textbf{data3})} \\
\noindent repeating up to \texttt{dataN} can be generalized to: \\
\noindent\texttt{(s3:getObject,arn:aws:s3:::category/\textbf{data*})} \\
Without generalization, the LLM may add a specific rule for each request, correctly classifying provided requests but failing on similar unseen ones.

\textbf{Setup.} For each policy, we selected one misclassified request and generated 10 resource variants (e.g., \texttt{data1} through \texttt{data10}), producing 20 total requests per policy. We repaired each policy using both approaches, then tested generalizability by evaluating the repaired policies on 15 held-out variant requests not provided during the repair process.

\begin{figure}[t]
  \centering
  \includegraphics[width=0.9\linewidth]{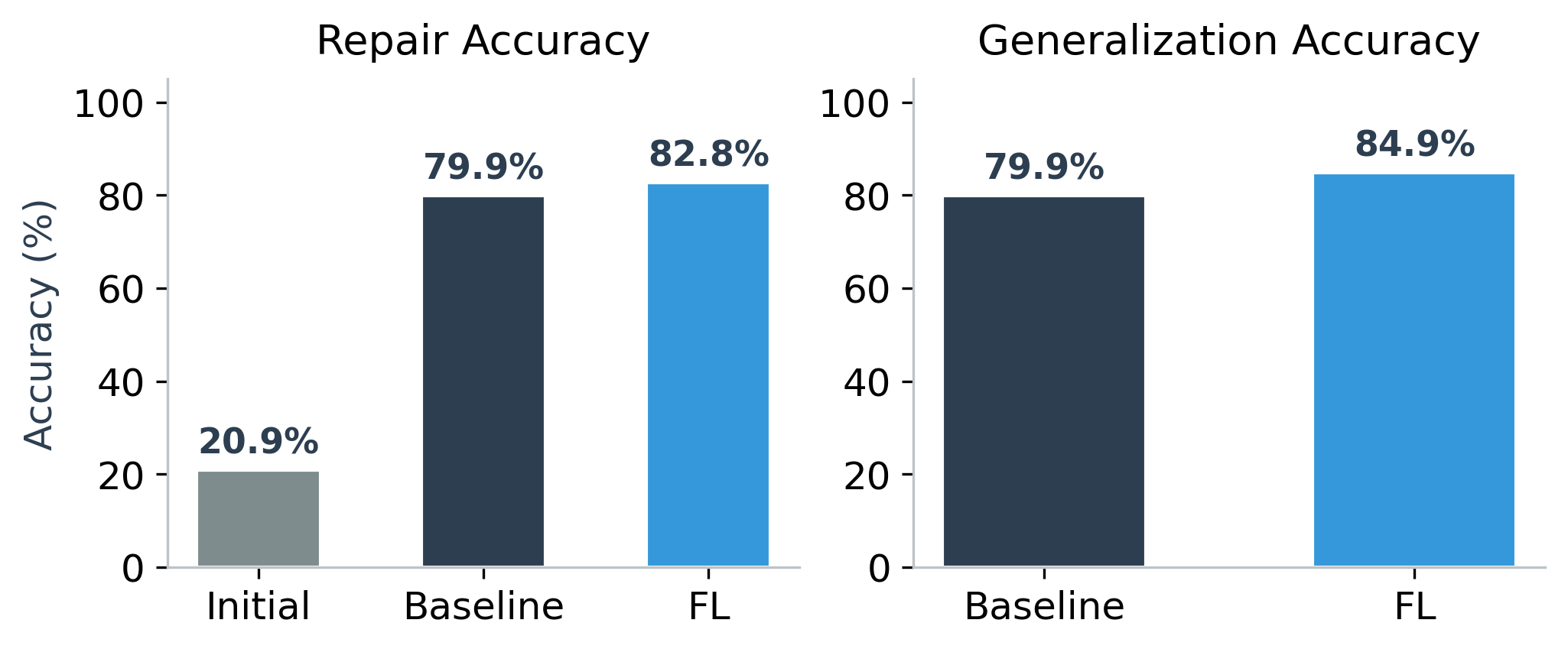}
\caption{Left panel: compares the repair accuracy of the baseline (79.9\%) and Fault Localization guided (82.8\%) methods against the initial faulty policy (20.9\%). Right panel: evaluates how these repairs generalize to requests not provided during the repair process, showing Fault Localization (84.9\%) can generate more generalizable repairs than the baseline (79.9\%).}
  \label{fig:generalization}
\end{figure}

\textbf{Results.} Figure~\ref{fig:generalization} presents the repair and generalization accuracy for both approaches. The left panel shows that Fault Localization guided repair achieves 82.8\% (172 complete repairs) accuracy compared to the baseline's 79.9\% (147 complete repairs), showing a 2.9\% improvement over the baseline. The right panel shows that the baseline achieves 79.9\% accuracy on unseen request variants, and Fault Localization achieves 84.9\%. Interestingly, the baseline generalization accuracy matches its repair accuracy (79.9\%). This suggests that when the baseline generates complete repairs, it produces repairs that generalize well to semantically similar requests. The Fault Localization approach shows stronger generalization, achieving 84.9\% on variants compared to its 82.8\% repair accuracy. 

\textbf{RQ5: Can CloudFix repair policies when requests are synthesized from natural language descriptions of user intent?}
Our final research question examines whether CloudFix can repair policies when an LLM from natural language descriptions of user intent generates requests.
In practice, administrators often know \textit{what} access they want to allow or deny but may lack request specifications to test their policies. Manually crafting such specifications is tedious and prone to errors. If LLMs can translate natural language intent into concrete requests, CloudFix becomes applicable even when requests are unavailable. \\
\textbf{Setup.} We select 10 faulty policies from our dataset. As mentioned in Section IV, our AWS IAM dataset contains 45 policies that have
\textit{faulty policy} (the original, incorrect policy posted by a user experiencing access control issues), \textit{ground truth policy} (the corrected policy provided by an AWS Cloud Engineer), and \textit{user intent} (the natural language description of the request specifications or problem as described by the user). We first use GPT-5, accessed via its web interface, to synthesize requests. We follow this procedure: for each policy, we prompted the model with the \textit{user intent}, \textit{faulty policy}, and \textit{ground truth policy} and instructed it to generate 5 to 10 valid requests. The criteria to generate a valid request were that they must be misclassified by the faulty policy, correctly classified by the ground truth, and encapsulate the user's intent. We then used Alg. \ref{alg:validate} to validate that the requests were misclassified by the faulty policy and correctly classified by the ground truth policy. To validate if the requests align with the user's expressed intent, we manually verified to ensure alignment. Next, we used \textit{CodeLlama-7B-Instruct} to repair the 10 faulty policies with a max of 5 iterations and a 100\% target accuracy.

\begin{figure}[ht]
  \centering
  \includegraphics[width=0.7\linewidth]{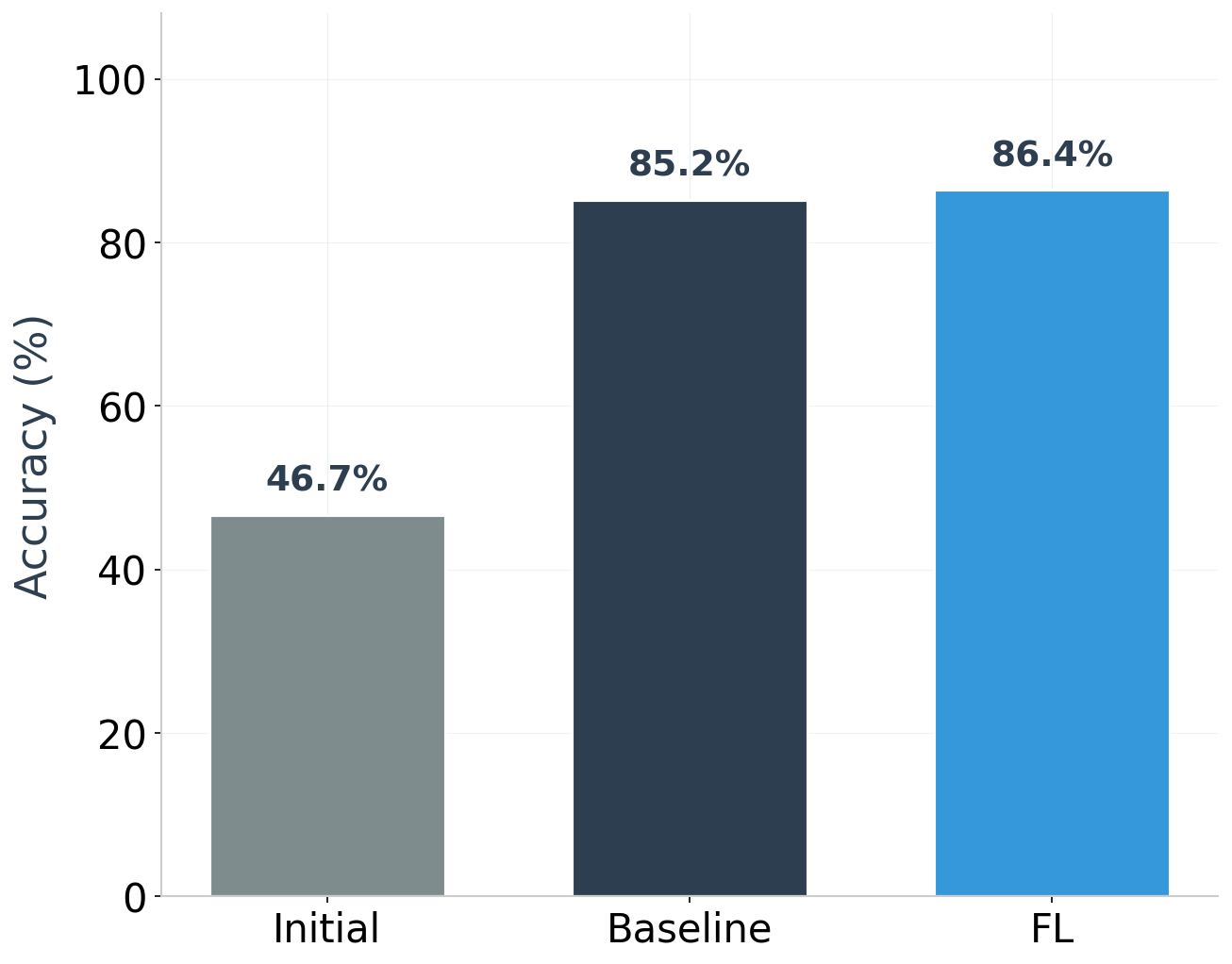}
  \caption{The results of CloudFix on requests generated from natural language descriptions. 
The marginal difference between the two methods suggests that LLM-generated requests from user intent effectively capture nuanced semantic differences.}
  \label{fig:intent}
\end{figure}

\textbf{Results.} Figure \ref{fig:intent} shows that the baseline approach achieves 85.2\% repair accuracy, while Fault Localization achieves 86.4\%, both showing improvements over the 46.7\% initial accuracy. This demonstrates that our framework can be used to generate policy repairs using requests generated by LLMs from natural language descriptions of the users' intents.

\section{Discussion and Threat to Validity}~\label{sec:discussion}
 We provide a qualitative analysis and discuss our findings. To understand the behavior of CloudFix at scale, we analyzed the failure and success modes of CodeLlama-7B-Instruct. The 7 successful repairs at request size 50 all involved single-statement policies amenable to straightforward additive repair, with 86\% completing within 2 iterations. The model consistently avoided overly permissive wildcards, preferring to enumerate specific requests. For moderate (80–99\%) repairs, the LLM generated syntactically valid policies but introduced semantic errors such as omitting explicit deny statements. For failed ($<$80\%) repairs, we observed more severe failures: condition omission (e.g., dropping \texttt{IpAddress} constraints), entity hallucination (substituting \texttt{arn:aws:iam::123456789012:root} for the actual principal), and syntactic degeneration where 46\% of policies showed no improvement due to repeated invalid JSON generation until token exhaustion. These behaviors align with known LLM tendencies to revert to high-frequency training patterns on out-of-distribution inputs~\cite{huang2025survey}.

Building on these observations, our main contribution is to show that fault localization guided prompting enables LLMs to repair policies with improved accuracy and time, particularly for moderately complex request sizes (RQ1, RQ2). The choice of LLM significantly impacts performance (RQ3), with CodeLlama-7B-Instruct achieving the highest complete repair rate, likely due to its code-focused pretraining providing familiarity with structured JSON, though we cannot isolate specific contributing factors. Results from RQ4 show that repairs generalize to unseen, semantically similar requests, and RQ5 demonstrates that LLM-synthesized requests from natural language can drive effective repair.

The baseline took significantly more time across all request sizes (Table~\ref{tab:timing_transposed}) as it enumerated requests individually, resulting in failure. We did not impose a timeout to ensure fair evaluation. Our dataset relies on policies from \href{https://repost.aws/questions}{AWS re:Post} with algorithmically generated requests, which may not fully capture real-world complexity. Additionally, generalizability is constrained by request size: accuracy degraded as request size grew to 50, with Fault Localization generating only 7 complete repairs.

Beyond accuracy, the practicality of CloudFix depends on computational efficiency. The Fault Localization guided approach is significantly more time-efficient than the baseline, though $>$99\% of repair time is spent on LLM inference. We restricted evaluation to smaller open-source models due to hardware constraints and to demonstrate that privacy-preserving policy repair is achievable without proprietary models. Our results show that combining post-hoc verification with Fault Localization provides formal correctness guarantees for small to medium request sizes while maintaining policy confidentiality. 

\section{Related Work}~\label{sec:related}
\textbf{Access Control Policy Repair}: Significant research has been conducted towards verifying and maintaining correct access control policies~\cite{10.5555/3161326,DetecingResolvingPolicyMisconfig,10.1145/1978942.1979243,KERN2022103301,Jayaraman2014AutomatedAA,Son2013FixMU,UsabilityChallengesAccessControl,UsabilityConsiderationsAccessControl,EffectsPolicyConflictResoution}. Numerous access control policy repair techniques based on formal methods have been proposed to fix policies to adhere to the principle of least permissiveness. Prior work by~\cite{xu2016towards} introduces a fault localization and mutation-based policy repair method to debug XACML access control policies. D'Antoni et al.~\cite{d2024automatically} proposed a synthesis-based framework to automatically reduce permissiveness in (AWS) IAM policies by leveraging access logs. Similarly,~\cite{peng2017towards} introduced fault localization and mutation-based policy repair techniques to produce a list of suspicious elements in XACML policies by correlating the test results and the test coverage information. They used mutation-based policy repair searches for fixes by mutating suspicious policy elements with predefined mutation operators. Eiers et al.~\cite{eiers2023quantitative} proposed a quantitative symbolic analysis framework to automatically repair overly permissive cloud access control policies. They encoded policy semantics as SMT formulas and used model counting to quantify and iteratively reduce the permissiveness of access control policies. However, their approach cannot repair semantically faulty policies. Additionally, traditional symbolic repair tools, such as those by~\cite{D'Antoni2024} and~\cite{eiers2022quantifying}, are limited to specific, hard-coded repair patterns and will not generalize to semantically complex policies. Conversely, our approach leverages the pattern recognition capabilities of LLMs to understand the semantic relationship among large and high-dimensional policies and requests to synthesize complex repairs beyond simple mutations.

\textbf{LLM-Based Automated Program Repair}: Recent methods have shown promising results that LLMs can be effectively applied to automated program repair
\cite{zhang2024systematic,li2025hybrid,hossain2024deep,xia2023automated,campos2025empirical}.
These approaches frame bug fixing as a code generation problem. A recent work closely related to ours, \cite{orvalho2025counterexample}, introduces a formal method based on counterexamples to guide LLMs to provide more accurate program repairs. However, to the best of our knowledge, there is no prior work that has explored automated access control policy repair with LLMs guided by a fault localization. 
\section{Conclusion and Future Work}~\label{sec:conclusion}
In this work, we introduced CloudFix, the first automated framework to integrate LLMs with Formal Methods for access control policy repair. Manually repairing policies is time-consuming and prone to errors. We address this problem by developing a framework to automatically create policy repairs with LLMs and by enhancing their reasoning capabilities with targeted repairs, enabled by Fault Localization guidance. 
Our evaluation of a curated dataset of 282 real-world AWS policies showed that LLMs guided by Fault Localization can be used to repair cloud access control policies with significantly improved accuracy compared to the baseline approach. Future work could explore larger open-source models with greater context capacity, and apply supervised fine-tuning or reinforcement learning with verifiable rewards—the low computational cost of the SMT verifier makes it a practical reward signal.

\section{Data Availability}\label{sec:data}
The policy dataset and accompanying source code used in this study are publicly available at \url{https://github.com/bethelhall/fixmypolicy}.

\bibliographystyle{IEEEtran}
\bibliography{references}

\end{document}